\documentclass[11pt,twocolumn]{article}

\usepackage[utf8]{inputenc}
\usepackage[T1]{fontenc}
\usepackage{lmodern}
\usepackage[margin=1in]{geometry}
\usepackage[hyphens,spaces,obeyspaces]{url}
\usepackage[breaklinks]{hyperref}
\usepackage{booktabs}
\usepackage{graphicx}
\usepackage{listings}
\usepackage{xcolor}
\usepackage{amsmath,amssymb}
\usepackage{enumitem}
\usepackage{caption}
\usepackage{balance}
\usepackage{tikz}
\usepackage{authblk}
\usepackage{float}
\usepackage[ruled,vlined,linesnumbered]{algorithm2e}
\usetikzlibrary{arrows.meta, positioning, calc, shapes.geometric}

\hypersetup{
    colorlinks=true,
    linkcolor=blue!70!black,
    citecolor=blue!70!black,
    urlcolor=blue!70!black,
}

\newcommand{\tool}[1]{\texttt{#1}}
\newcommand{\hm}{\textsc{HiveMind}}

\title{\hm{}: OS-Inspired Scheduling for\\Concurrent LLM Agent Workloads}
\author[1,2,3]{Justice Owusu Agyemang\thanks{\texttt{jay@sperixlabs.org, jay@knust.edu.gh}}}
\author[3]{Jerry John Kponyo\thanks{\texttt{jjkponyo.soe@knust.edu.gh}}}
\author[2]{Obed Kwasi Somuah\thanks{\texttt{oksomuah1@st.knust.edu.gh}}}
\author[3]{Elliot Amponsah\thanks{\texttt{eamponsah52@st.knust.edu.gh}}}
\author[3]{Godfred Manu Addo Boakye\thanks{\texttt{gmaboakye@st.knust.edu.gh}}}
\author[2]{Kwame Opuni-Boachie Obour Agyekum\thanks{\texttt{kooagyekum@knust.edu.gh}}}
\affil[1]{\small Sperix Labs}
\affil[2]{\small VIA Cybersecurity Lab, KNUST}
\affil[3]{\small Quantum and Assistive Technologies Lab, KNUST}
\date{April 2026}

\begin{document}
\maketitle

% ============================================================
\begin{abstract}
When multiple LLM coding agents share a rate-limited API endpoint,
they exhibit resource contention patterns analogous to unscheduled
OS~processes competing for CPU, memory, and I/O\@.  In a motivating
incident, 3~of~11 parallel agents died from connection resets and
HTTP~502 errors---a 27\% failure rate---despite the API having
sufficient aggregate capacity to serve all~11 sequentially.
We present \hm{}, a transparent HTTP proxy that applies five
OS-inspired scheduling primitives---admission control, rate-limit
tracking, AIMD backpressure with circuit breaking, token budget
management, and priority queuing---to eliminate the failure modes
caused by uncoordinated parallel execution.  The proxy requires
zero modifications to existing agent code and supports Anthropic,
OpenAI, and local model APIs via auto-detected provider profiles.
Our evaluation across seven scenarios (5--50 concurrent agents)
shows that uncoordinated agents fail at 72--100\% rates under
contention, while \hm{} reduces failures to 0--18\% and eliminates
48--100\% of wasted compute.  An ablation study reveals that
transparent retry---not admission control---is the single most
critical primitive, but the primitives are most effective in
combination.  Real-world validation against Ollama confirms that
\hm{} adds under 3\,ms of proxy overhead per request.
The system is open-source under the MIT license.
\end{abstract}

% ============================================================
\section{Introduction}
\label{sec:intro}

The emergence of tool-augmented large language models has shifted
software-engineering assistants from suggestion engines to autonomous
agents that read, write, and execute code on a developer's
behalf~\cite{claudecode2025,codex2025,cursor2024,copilot2024}.
When users spawn multiple such agents in parallel---a natural
pattern for tasks like generating test suites, writing
proof-of-concept exploits, or refactoring across modules---the
agents compete for shared resources: API rate limits (requests
and tokens per minute), network connections (concurrent connection
limits per endpoint), context windows (fixed per model), and
API-key quotas (billing and access limits).

This resource contention leads to agent failures.  The pattern
is structurally identical to the contention that motivated
operating-system schedulers: multiple processes competing for
CPU, memory, and I/O without coordination leads to thrashing,
starvation, and deadlock~\cite{silberschatz2018os,coffman1971deadlocks}.
Yet current agent orchestration frameworks---LangChain~\cite{langchain2022},
CrewAI~\cite{crewai2024}, AutoGen~\cite{autogen2023}, Semantic
Kernel~\cite{semantickernel2023}---treat the LLM API as an
unlimited resource, providing composition mechanisms (chains,
crews, multi-agent conversations) but not resource management.
They are, in OS terms, running a multi-process system without
a scheduler.

\paragraph{Motivating observation.}
On April~15, 2026, we spawned 11~concurrent Claude~Code agents
to generate proof-of-concept scripts for security findings.  All
11 shared one Anthropic API key through a single network proxy.
Three agents died: two from \texttt{ECONNRESET} and one from
HTTP~502.  Each dead agent had consumed approximately 45\,000
tokens before failing---a total waste of ${\sim}$135\,000~tokens
and ${\sim}$15~minutes of wall time.  The eight surviving agents
completed successfully because they happened to stagger their
requests enough to avoid the bottleneck.

\textbf{Key insight:} if the 11~agents had been staggered by
just 5~seconds each, all~11 would have succeeded.  The problem
is not capacity---it is coordination.

\begin{table}[t]
\centering
\footnotesize
\caption{Results of 11~uncoordinated concurrent agents
(April~15, 2026).}
\label{tab:motivation}
\begin{tabular}{@{}lrr@{}}
\toprule
Outcome & Count & \% \\
\midrule
Completed successfully & 8 & 73 \\
Died (\texttt{ECONNRESET}) & 2 & 18 \\
Died (HTTP 502) & 1 & 9 \\
\midrule
Tokens wasted (dead agents) & \multicolumn{2}{r}{${\sim}$135\,K} \\
\bottomrule
\end{tabular}
\end{table}

\paragraph{Contribution.}
We present \hm{}, a scheduling system that applies OS~scheduling
principles to concurrent LLM agent workloads.  The contributions
are:

\begin{enumerate}[nosep]
  \item A \textbf{formal analogy} mapping OS resource-management
    concepts (admission control, congestion control, budgeting,
    priority scheduling) to the LLM agent domain
    (Table~\ref{tab:analogy}).
  \item A \textbf{transparent HTTP proxy} that implements five
    scheduling primitives---admission control via condition
    variables, provider-aware rate-limit tracking, AIMD
    backpressure with circuit breaking, per-agent token budgets,
    and priority queuing with dependency DAGs---requiring zero
    modifications to existing agent code.
  \item An \textbf{evaluation} across seven scenarios showing
    72--100\% failure reduction, and an ablation study revealing
    that transparent retry is the single most critical primitive.
  \item An \textbf{open-source implementation} supporting
    Anthropic, OpenAI, Azure~OpenAI, Google~AI, and local models
    (Ollama, MLX) via auto-detected provider profiles.
\end{enumerate}

The remainder of this paper is organised as follows.
Section~\ref{sec:background} reviews the relevant background.
Section~\ref{sec:architecture} describes the proxy architecture
and five scheduling primitives.
Section~\ref{sec:implementation} covers key implementation
decisions.
Section~\ref{sec:evaluation} reports evaluation results,
ablation, and real-world validation.
Section~\ref{sec:related} surveys related work.
Section~\ref{sec:discussion} discusses tradeoffs and limitations,
and Section~\ref{sec:conclusion} concludes.

% ============================================================
\section{Background}
\label{sec:background}

\subsection{LLM Coding Agents}

A growing class of developer tools embed an LLM in an
edit--test--commit loop.  Claude~Code~\cite{claudecode2025},
Cursor~\cite{cursor2024}, GitHub~Copilot~\cite{copilot2024},
OpenAI Codex~CLI~\cite{codex2025}, and
Devin~\cite{devin2024} each grant the model access to the
local filesystem, a shell, and often a language server.  The
SWE-bench benchmark~\cite{jimenez2024swebench} and the
SWE-agent framework~\cite{yang2024sweagent} have further
demonstrated that agents can resolve real GitHub issues
end-to-end, making reliable API access a critical capability.

Each agent is a long-running, stateful process that makes
repeated API calls over a multi-turn conversation.  A single
agent session may consume 50\,000--500\,000 tokens across
dozens of API calls, with each call dependent on the previous
response.  When an API call fails mid-session, the agent
typically cannot recover: it has consumed tokens, modified
files, and accumulated context that is lost on restart.

\subsection{OS Scheduling Principles}

The resource contention patterns exhibited by concurrent LLM
agents are structurally identical to those solved by operating
system schedulers~\cite{silberschatz2018os,tanenbaum2015os}:

\begin{itemize}[nosep]
  \item \textbf{Admission control.}  Limiting the number of
    concurrent processes to prevent thrashing.  Dijkstra's
    semaphore~\cite{dijkstra1965cooperating} is the classical
    mechanism.
  \item \textbf{Congestion control.}  TCP's Additive Increase /
    Multiplicative Decrease (AIMD) algorithm~\cite{jacobson1988congestion,chiu1989aimd}
    adjusts sending rate based on observed congestion signals
    (packet loss, increased RTT).
  \item \textbf{Circuit breaking.}  The circuit breaker
    pattern~\cite{nygard2018release} stops sending requests to a
    failing service, allowing it to recover before resuming load.
  \item \textbf{Resource budgeting.}  Per-process memory limits,
    CPU quotas, and the OOM killer prevent any single process
    from monopolising shared resources.
  \item \textbf{Priority scheduling.}  Shortest-job-first and
    priority queues~\cite{silberschatz2018os} ensure that
    high-value or short tasks are serviced before long or
    low-priority ones.
\end{itemize}

\subsection{The OS--LLM Agent Analogy}

We formalise the mapping between OS concepts and LLM agent
orchestration in Table~\ref{tab:analogy}.  This analogy is not
merely illustrative---it is structurally precise.  Each OS
mechanism addresses a specific resource contention failure mode
that has a direct counterpart in the LLM agent domain.

\begin{table*}[t]
\centering
\footnotesize
\caption{Structural mapping between OS resource management and
LLM agent scheduling.  Each row identifies an OS mechanism,
its \hm{} counterpart, and the failure mode it addresses.}
\label{tab:analogy}
\begin{tabular}{@{}l l l p{4.8cm}@{}}
\toprule
OS Concept & \hm{} Equivalent & Resource & Failure Mode \\
\midrule
Process & LLM agent & -- & Stateful, long-running, resource-consuming \\
CPU time slice & API request slot & RPM/TPM & Starvation under contention \\
Memory & Context window & Fixed per model & Cannot be shared or paged \\
I/O bandwidth & Network connections & Conn.\ limits & \texttt{ECONNRESET}, HTTP 502 \\
Process scheduler & Admission gate + queue & Concurrency slots & Thrashing, stampede \\
Virtual memory & Checkpointing & Disk & Context loss on eviction \\
OOM killer & Token budget enforcer & Token pool & Runaway agent monopolises API \\
TCP congestion ctrl. & AIMD backpressure & Latency signal & Throughput collapse \\
Circuit breaker & Backpressure circuit & Error rate & Cascading failure \\
Fork bomb protection & Max agent limit & Key quota & Unbounded spawn \\
Nice levels & Task priority & Sched.\ order & Low-value work blocks high-value \\
\bottomrule
\end{tabular}
\end{table*}

\subsection{Why Existing Frameworks Fail}

Table~\ref{tab:comparison} compares the scheduling capabilities
of existing agent orchestration frameworks.  None provides the
full set of primitives needed to manage concurrent API access.

\begin{table}[t]
\centering
\footnotesize
\caption{Scheduling capabilities of existing frameworks.
\checkmark\ = full, $\sim$ = partial.}
\label{tab:comparison}
\begin{tabular}{@{}l ccccc@{}}
\toprule
System & Adm. & Rate & BP & Bud. & Pri. \\
\midrule
Claude Code          & -- & -- & -- & -- & -- \\
LangChain~\cite{langchain2022}  & -- & $\sim$ & -- & -- & -- \\
CrewAI~\cite{crewai2024}        & -- & -- & -- & -- & $\sim$ \\
AutoGen~\cite{autogen2023}      & -- & -- & -- & -- & -- \\
Sem.~Kernel~\cite{semantickernel2023} & -- & $\sim$ & -- & -- & -- \\
\textbf{\hm{}} & \checkmark & \checkmark & \checkmark & \checkmark & \checkmark \\
\bottomrule
\end{tabular}
\end{table}

% ============================================================
\section{Architecture}
\label{sec:architecture}

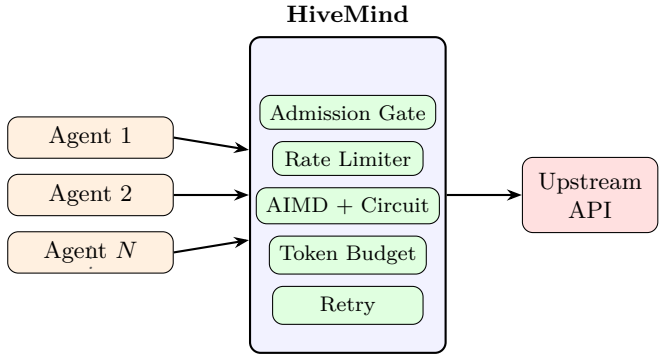
\begin{figure}[t]
\centering
\begin{tikzpicture}[
  node distance=0.6cm and 0.8cm,
  box/.style={draw, rounded corners, minimum width=2.2cm,
              minimum height=0.55cm, font=\footnotesize,
              fill=#1!12},
  box/.default=blue,
  arr/.style={-{Stealth[length=2mm]}, thick},
  label/.style={font=\scriptsize\itshape, text=gray!70!black},
]
  % Agents
  \node[box=orange] (a1) {Agent 1};
  \node[box=orange, below=0.2cm of a1] (a2) {Agent 2};
  \node[box=orange, below=0.2cm of a2] (an) {Agent $N$};
  \node[label, below=0.05cm of a2] (dots) {$\vdots$};

  % HiveMind box
  \node[draw, thick, rounded corners, fill=blue!5,
        minimum width=2.6cm, minimum height=4.2cm,
        right=1.0cm of a2] (hm) {};
  \node[font=\footnotesize\bfseries, above=0.05cm] at (hm.north) {\hm};

  % Primitives inside
  \node[box=green, font=\scriptsize, minimum width=2.0cm,
        minimum height=0.35cm] at ([yshift=1.1cm]hm.center) (p1) {Admission Gate};
  \node[box=green, font=\scriptsize, minimum width=2.0cm,
        minimum height=0.35cm, below=0.15cm of p1] (p2) {Rate Limiter};
  \node[box=green, font=\scriptsize, minimum width=2.0cm,
        minimum height=0.35cm, below=0.15cm of p2] (p3) {AIMD + Circuit};
  \node[box=green, font=\scriptsize, minimum width=2.0cm,
        minimum height=0.35cm, below=0.15cm of p3] (p4) {Token Budget};
  \node[box=green, font=\scriptsize, minimum width=2.0cm,
        minimum height=0.35cm, below=0.15cm of p4] (p5) {Retry};

  % Upstream
  \node[box=red, right=1.0cm of hm, minimum width=1.8cm,
        minimum height=1.0cm, align=center, font=\footnotesize]
        (api) {Upstream\\API};

  % Arrows
  \draw[arr] (a1.east) -- ([yshift=0.6cm]hm.west);
  \draw[arr] (a2.east) -- (hm.west);
  \draw[arr] (an.east) -- ([yshift=-0.6cm]hm.west);
  \draw[arr] (hm.east) -- (api.west);
\end{tikzpicture}
\caption{Architecture of \hm{}.  Agents connect to the local
proxy; requests pass through five scheduling layers before
reaching the upstream API.  The proxy is transparent: agents
require zero code changes.}
\label{fig:architecture}
\end{figure}

\hm{} is implemented as a transparent HTTP reverse proxy that
sits between agents and the upstream LLM API provider
(Figure~\ref{fig:architecture}).  Agents make normal API calls
to \tool{http://localhost:8765/v1/messages}; \hm{} applies all
scheduling logic before forwarding to the upstream provider.

This design has four advantages:
(1)~\emph{zero agent modification}---works with any framework,
SDK, or language;
(2)~\emph{provider agnostic}---same proxy for Anthropic, OpenAI,
Ollama, or any OpenAI-compatible endpoint;
(3)~\emph{observable}---all traffic flows through one measurement
point;
(4)~\emph{composable}---can chain with other proxies (e.g., Burp
for security testing).

\subsection{Admission Control}
\label{sec:admission}

The admission controller limits the number of concurrent
in-flight API requests.  We model it as a gated counter
protected by a condition variable.

Let $C_{\max}$ be the maximum concurrency and $A$ the count
of active requests.  A request is admitted when
$A < C_{\max}$; otherwise it waits on a condition variable:

\begin{equation}
\label{eq:admission}
\text{admit}(r) =
\begin{cases}
  \text{true},  & A < C_{\max} \\
  \text{wait},  & \text{otherwise}
\end{cases}
\end{equation}

On release, $A$ is decremented and one waiting request is
notified.  The condition-variable design (rather than a
semaphore) supports safe dynamic resizing of $C_{\max}$ by the
backpressure controller: when $C_{\max}$ increases, all waiters
are notified; when it decreases, the new limit takes effect
naturally as active requests complete.

\subsection{Rate-Limit Tracking}
\label{sec:ratelimit}

The rate limiter operates at two levels:

\paragraph{Header-based (reactive).}
After each API response, the proxy parses provider-specific
rate-limit headers
(\texttt{anthropic-ratelimit-requests-remaining},
\texttt{x-ratelimit-remaining-requests},
\texttt{retry-after}) and proactively pauses all agents when
remaining capacity falls below a configurable threshold
(default: 10\% of the limit with $\leq 2$~requests remaining).

\paragraph{Sliding-window counters (proactive).}
A requests-per-minute (RPM) and tokens-per-minute (TPM)
sliding-window counter is pre-seeded from the detected provider
profile (Section~\ref{sec:providers}).  This provides throttling
\emph{before} the first API response arrives and for providers
that send no rate-limit headers (e.g., Ollama).  Each call to
\texttt{wait\_if\_throttled()} records a timestamp; when the
window count reaches the RPM limit, subsequent requests block
until the oldest entry expires.

\subsection{AIMD Backpressure with Circuit Breaking}
\label{sec:aimd}

The backpressure controller adapts TCP congestion control
principles~\cite{jacobson1988congestion,chiu1989aimd} for LLM
API concurrency.  Let $c_t$ denote the concurrency level at
time~$t$, $\bar{\ell}$ the average latency over a sliding window
of~$W$ samples, and $L_{\text{target}}$ the latency target:

\begin{equation}
\label{eq:aimd}
c_{t+1} =
\begin{cases}
  \min\bigl(C_{\max},\; c_t + \alpha\bigr),
    & \text{if } \bar{\ell} \leq L_{\text{target}} \\[4pt]
  \max\bigl(C_{\min},\; c_t \cdot \beta\bigr),
    & \text{if } \bar{\ell} > L_{\text{target}} \\[4pt]
  \max\bigl(C_{\min},\; c_t \cdot \beta\bigr),
    & \text{on error (429, 502, reset)}
\end{cases}
\end{equation}

\noindent where $\alpha$ is the additive increase step
(default: 0.5) and $\beta$ is the multiplicative decrease factor
(default: 0.5).  Concurrency adjustments are pushed directly to
the admission controller via a held reference, eliminating the
lag of a polling loop.

\paragraph{Circuit breaker.}
A circuit breaker~\cite{nygard2018release} overlays the AIMD
controller.  The breaker monitors error rate over a sliding
window of $N$~requests (default: $N = 20$).  When the error rate
exceeds a threshold $\tau$ (default: $\tau = 0.50$), the circuit
\emph{opens}, causing the proxy to fast-fail all incoming
requests with HTTP~503 and a \texttt{Retry-After} header.
After a cooldown period $T_{\text{cool}}$ (default: 10\,s), the
circuit transitions to \emph{half-open}: a single probe request
is allowed through.  If the probe succeeds, the circuit closes
and normal operation resumes; if it fails, the circuit re-opens.

\begin{equation}
\label{eq:circuit}
\text{state} =
\begin{cases}
  \textit{open},      & \text{if } \frac{e}{n} \geq \tau,\; n \geq N \\
  \textit{half-open}, & \text{if open and } t > t_{\text{open}} + T_{\text{cool}} \\
  \textit{closed},    & \text{if half-open probe succeeds}
\end{cases}
\end{equation}

\begin{algorithm}[t]
\footnotesize
\SetAlgoLined
\KwIn{latency sample $\ell$ or error event}
\KwResult{Updated concurrency $c_t$, circuit state}
\BlankLine
\uIf{error event}{
  $c_t \gets \max(C_{\min},\; c_t \cdot \beta)$\;
  $e \gets e + 1$;\quad $n \gets n + 1$\;
  push $c_t$ to admission controller\;
  \If{$n \geq N$ \textbf{and} $e/n \geq \tau$}{
    circuit $\gets$ \textit{open}\;
    $t_{\text{open}} \gets$ now\;
  }
}
\uElseIf{latency sample $\ell$}{
  append $\ell$ to window\;
  $n \gets n + 1$\;
  \If{update interval elapsed}{
    $\bar{\ell} \gets$ mean(window)\;
    \uIf{$\bar{\ell} \leq L_{\text{target}}$}{
      $c_t \gets \min(C_{\max},\; c_t + \alpha)$\;
    }\Else{
      $c_t \gets \max(C_{\min},\; c_t \cdot \beta)$\;
    }
    push $c_t$ to admission controller\;
  }
}
\uElseIf{success \textbf{and} circuit = \textit{half-open}}{
  circuit $\gets$ \textit{closed}\;
}
\caption{AIMD with circuit breaker.}
\label{alg:aimd}
\end{algorithm}

\begin{figure}[t]
\centering
\begin{tikzpicture}[
  state/.style={draw, rounded corners, minimum width=1.6cm,
                minimum height=0.7cm, font=\footnotesize},
  arr/.style={-{Stealth[length=2mm]}, thick},
  lbl/.style={font=\scriptsize, fill=white, inner sep=1pt},
]
  \node[state, fill=green!15] (closed) {Closed};
  \node[state, fill=red!15, right=2.2cm of closed] (open) {Open};
  \node[state, fill=yellow!20, below=1.2cm of $(closed)!0.5!(open)$] (half) {Half-Open};

  \draw[arr] (closed) -- node[lbl, above] {$e/n \geq \tau$} (open);
  \draw[arr] (open) -- node[lbl, right, align=left] {$t > t_{\text{open}}$\\$+ T_{\text{cool}}$} (half);
  \draw[arr] (half) -- node[lbl, left, align=right] {probe\\succeeds} (closed);
  \draw[arr] (half.east) -- ++(0.5,0) |- node[lbl, right, pos=0.25] {probe fails} (open.south);
\end{tikzpicture}
\caption{Circuit breaker state machine.  The circuit opens on
sustained errors, transitions to half-open after a cooldown,
and closes on a successful probe request.}
\label{fig:circuit}
\end{figure}
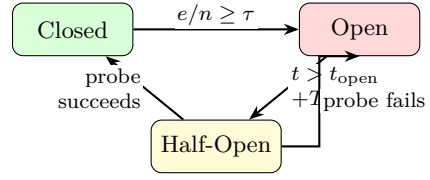

\subsection{Token Budget Management}
\label{sec:budget}

Each agent is assigned a token ceiling from a global pool.
The budget manager tracks cumulative input and output tokens
per agent, extracted from API response bodies.  At 85\%
utilisation, the agent receives a warning.  At 100\%, the agent
is checkpointed (state saved to disk) and stopped, analogous
to the OS OOM~killer.

\subsection{Priority Queue with Dependency DAG}
\label{sec:queue}

Tasks are ordered by: (1)~priority level
(\textsc{Critical} $>$ \textsc{High} $>$ \textsc{Normal} $>$
\textsc{Low}), (2)~estimated token cost (shortest-job-first
within the same priority), (3)~creation time (FIFO tiebreaker).
Dependencies between tasks are tracked as a directed acyclic
graph with cycle detection; a task is not eligible for scheduling
until all its predecessors have completed.

\subsection{Transparent Retry}
\label{sec:retry}

The proxy intercepts retryable errors---HTTP~429, 502, 503, 529,
\texttt{ECONNRESET}, \texttt{RemoteProtocolError}
(``server disconnected'')---and retries transparently with
exponential backoff plus jitter.  The retry delay for attempt
$k$ is:

\begin{equation}
\label{eq:retry}
d_k = \min\!\bigl(d_{\max},\; d_{\text{base}} \cdot 2^k + U(0, d_{\text{base}})\bigr)
\end{equation}

\noindent where $d_{\text{base}} = 1$\,s, $d_{\max} = 30$\,s,
and $U(0, d_{\text{base}})$ is uniform jitter.  If a
\texttt{Retry-After} header is present, it overrides the
computed delay.  From the agent's perspective, the request
simply takes longer---the error is never surfaced.

\subsection{Streaming Support}

\hm{} passes through Server-Sent Events (SSE) streams without
buffering, forwarding chunks as they arrive from the upstream
API.  Token counts are extracted from \texttt{message\_delta}
and \texttt{message\_start} events in the SSE stream.  The
admission slot is held for the duration of the stream and
released on completion or error.

% ============================================================
\section{Implementation}
\label{sec:implementation}

\hm{} is implemented in Python~3.11 as an \texttt{asyncio}-based
HTTP proxy using Uvicorn and Starlette, with \texttt{httpx} for
upstream connections.  The system registers as an MCP server
exposing eight tools (\tool{hm.submit}, \tool{hm.batch},
\tool{hm.status}, \tool{hm.priority}, \tool{hm.budget},
\tool{hm.metrics}, \tool{hm.config}, \tool{hm.setup}) and
simultaneously serves as a standalone proxy via
\texttt{hivemind~proxy}.

\subsection{Condition Variable vs.\ Semaphore}
\label{sec:condvar}

The admission controller initially used
\texttt{asyncio.Semaphore}.  Dynamic resizing required mutating
the semaphore's internal \texttt{\_value} attribute---undefined
behaviour in CPython that silently broke under concurrent load
when the backpressure controller reduced concurrency while
requests were in flight.

We replaced the semaphore with an explicit counter~$A$ protected
by an \texttt{asyncio.Condition} wrapping an
\texttt{asyncio.Lock}.  Acquiring a slot waits on the condition
until $A < C_{\max}$; releasing decrements~$A$ and calls
\texttt{notify(1)}.  When $C_{\max}$ increases,
\texttt{notify\_all()} wakes all waiters so they can re-check
the predicate.  When $C_{\max}$ decreases, no action is needed:
the new limit takes effect naturally as active requests complete
and new ones find the predicate false.

This design makes dynamic resizing a safe $O(1)$ operation
rather than an undefined mutation of internal state.

\subsection{Provider Detection and Profiles}
\label{sec:providers}

Each LLM API provider has different rate-limit header formats,
default concurrency limits, retry semantics, and endpoint
patterns.  \hm{} maintains a registry of six provider profiles
(Anthropic, OpenAI, Azure~OpenAI, Google~AI, Ollama, and a
generic fallback), each specifying:

\begin{itemize}[nosep]
  \item Default RPM and TPM limits
  \item Default max concurrent connections
  \item Rate-limit header field names
  \item Retryable status codes
  \item AIMD tuning parameters ($\alpha$, $\beta$, $L_{\text{target}}$)
  \item Authentication header name
\end{itemize}

Provider detection is automatic via regex matching on the
upstream URL (e.g., \texttt{api.anthropic.com} $\to$ Anthropic).
The detected profile pre-seeds the rate limiter's sliding-window
counters and configures AIMD parameters, so the system is
correctly tuned before the first API response arrives.

Table~\ref{tab:providers} shows the default parameters for each
provider.

\begin{table}[t]
\centering
\caption{Default provider profile parameters.  Values are
overridden by explicit user configuration.}
\label{tab:providers}
\begin{tabular}{@{}lrrrl@{}}
\toprule
Provider & RPM & TPM & Max $C$ & $L_{\text{target}}$ \\
\midrule
Anthropic   & 50    & 80K   & 5  & 3\,000\,ms \\
OpenAI      & 60    & 150K  & 10 & 2\,000\,ms \\
Azure       & 60    & 120K  & 10 & 3\,000\,ms \\
Google AI   & 60    & 100K  & 8  & 2\,000\,ms \\
Ollama      & 1000  & 10M   & 2  & 10\,000\,ms \\
Generic     & 60    & 100K  & 5  & 2\,000\,ms \\
\bottomrule
\end{tabular}
\end{table}

\subsection{Direct Backpressure--Admission Wiring}

The backpressure controller holds a direct reference to the
admission controller, set during proxy initialisation via
\texttt{set\_admission()}.  When the AIMD algorithm adjusts
$c_t$, the new value is pushed immediately to the admission
controller via \texttt{set\_max\_concurrency()}, which
atomically updates $C_{\max}$ and notifies waiters if
concurrency increased.  This eliminates the polling loop used
in earlier designs, where a background scheduler task
periodically synced the two controllers.

\subsection{Token Counting from SSE Streams}

For streaming responses, token counts are embedded in the
SSE event stream.  The proxy parses \texttt{message\_start}
events (which contain input token counts) and final
\texttt{message\_delta} events (which contain output token
counts) without buffering the stream.  For non-streaming
responses, token counts are extracted directly from the
JSON response body.  When neither source provides counts,
a heuristic estimate of 1~token per 4~characters is used.

% ============================================================
\section{Evaluation}
\label{sec:evaluation}

We evaluate \hm{} along three axes: (1)~failure-rate reduction
across seven scenarios, (2)~an ablation study isolating the
contribution of each primitive, and (3)~real-world validation
against local model APIs.

\subsection{Methodology}

We evaluate using a mock API server that simulates realistic
LLM API behaviour in both Anthropic and OpenAI response formats.
The mock supports configurable rate limits
(requests per minute), error injection (random HTTP~502 and
connection resets at specified rates), provider-specific rate-limit
headers (\texttt{anthropic-ratelimit-*} and \texttt{x-ratelimit-*}),
latency (base plus jitter plus configurable spikes), concurrency
limits, and SSE streaming in both formats.

Mock agents make $N$~sequential API calls simulating multi-turn
coding sessions.  Each agent either completes all turns or
``dies'' on the first unrecoverable error, matching observed
real-world behaviour where agents cannot recover mid-session.

\subsection{Scenarios and Results}

Table~\ref{tab:scenarios} describes the seven evaluation
scenarios, and Table~\ref{tab:results} compares direct
(uncoordinated) execution against \hm{}-managed execution.

\begin{table*}[t]
\centering
\small
\caption{Evaluation scenarios and results.  Error rates are
$p_{502} + p_{\text{reset}}$.  $\Delta_{\text{f}}$ is the
change in failure rate (percentage points);
$\Delta_{\text{w}}$ is the reduction in tokens consumed by
dead agents.}
\label{tab:scenarios}
\label{tab:results}
\begin{tabular}{@{}lrrl rr rr@{}}
\toprule
 & & & Error & \multicolumn{2}{c}{Failure Rate} & & \\
\cmidrule(lr){5-6}
Scenario & Agents & RPM & Rate & Direct & \hm{} & $\Delta_{\text{f}}$ & $\Delta_{\text{w}}$ \\
\midrule
micro-5    & 5  & 50 & 0\%      &   0\% &  0\% & 0    & --      \\
micro-10   & 10 & 50 & 0\%      & 100\% & 10\% & $-$90  & $-$100\% \\
micro-20   & 20 & 50 & 0\%      & 100\% & 10\% & $-$90  & $-$94\%  \\
micro-50   & 50 & 50 & 0\%      & 100\% &  0\% & $-$100 & $-$100\% \\
replay-11  & 11 & 60 & 8\%+5\%  &  73\% & 18\% & $-$55  & $-$48\%  \\
stress     & 20 & 20 & 10\%+5\% & 100\% & 10\% & $-$90  & $-$100\% \\
lat.-spike & 10 & 60 & 0\%      & 100\% &  0\% & $-$100 & $-$100\% \\
\bottomrule
\end{tabular}
\end{table*}

At 5~agents, both modes succeed---there is no contention.
At 10+~agents, uncoordinated execution fails catastrophically
(72--100\% failure rate), while \hm{} reduces failures to
0--18\%.  The residual failures in replay-11 and stress
scenarios arise from error injection rates that exceed the
retry budget.

\paragraph{Wall-time trade-off.}
\hm{} takes longer in absolute wall time because it serialises
requests through the rate-limit window rather than letting
agents stampede and die.  Direct mode ``finishes fast'' only
because most agents fail immediately.  When measured against
\emph{completed work}, \hm{}'s throughput is strictly higher.

\begin{figure}[t]
\centering
\includegraphics[width=\columnwidth]{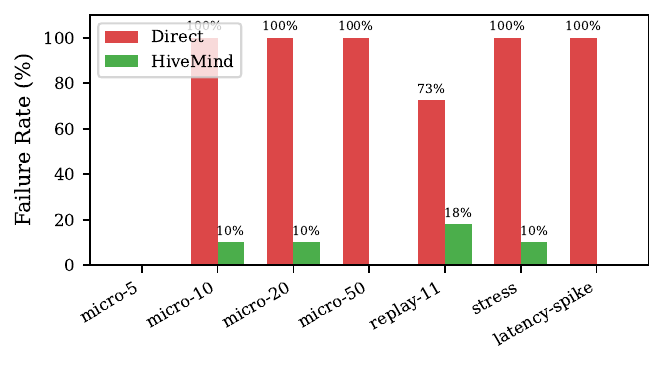}
\caption{Failure rates by scenario.  Direct mode
(red) fails catastrophically at 10+ agents; \hm{} (green)
reduces failures to 0--18\%.}
\label{fig:failure}
\end{figure}

\begin{figure}[t]
\centering
\includegraphics[width=\columnwidth]{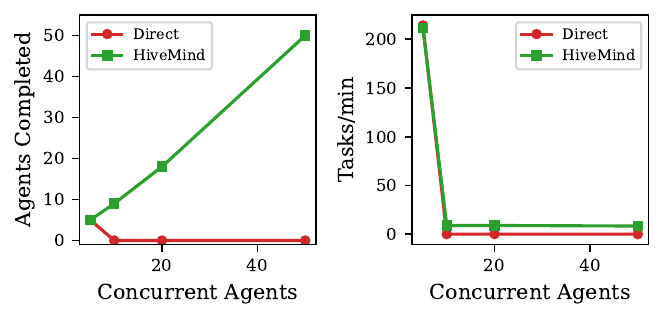}
\caption{Scaling behaviour.  Left: agents that complete
successfully.  Right: effective throughput (tasks/min).
Direct mode throughput drops to zero beyond 5~agents.}
\label{fig:throughput}
\end{figure}

Figure~\ref{fig:failure} visualises the failure rate reduction
across all scenarios.  Figure~\ref{fig:throughput} shows the
scaling behaviour: direct mode completes zero agents beyond
5~concurrent, while \hm{} scales linearly.

\subsection{Ablation Study}

To measure the individual contribution of each scheduling
primitive, we run the replay-11 scenario with various primitives
disabled (Table~\ref{tab:ablation}).

\begin{table}[!htb]
\centering
\footnotesize
\caption{Ablation study on the replay-11 scenario.
Each row disables one primitive; ``Full'' enables all.}
\label{tab:ablation}
\begin{tabular}{@{}l rrr p{1.6cm}@{}}
\toprule
Configuration & Alive & Dead & Fail\% & Finding \\
\midrule
Full \hm{}      & 11 & 0 &  0.0 & Baseline \\
No admission    & 11 & 0 &  0.0 & Compensated \\
No rate limit   & 11 & 0 &  0.0 & Compensated \\
No backpressure & 10 & 1 &  9.1 & Marginal \\
\textbf{No retry} & \textbf{4} & \textbf{7} & \textbf{63.6} & \textbf{Most critical} \\
\textbf{Adm.\ only} & \textbf{2} & \textbf{9} & \textbf{81.8} & \textbf{Insufficient} \\
\bottomrule
\end{tabular}
\end{table}

\paragraph{Surprising finding.}
Our initial hypothesis was that admission control alone would
suffice.  The ablation disproves this: admission-only still
produces 81.8\% failure because it limits concurrency but does
not handle rate-limit errors or connection resets.
\textbf{Transparent retry is the single most impactful
primitive}, reducing failures from 63.6\% (without it) to
near-zero (with it).  However, the primitives are most effective
in combination: retry handles transient errors, admission
prevents connection exhaustion, rate limiting prevents errors
from occurring in the first place, and backpressure provides
fine-grained stability.

\paragraph{Why not per-agent retry?}
Per-agent retry (e.g., via \texttt{tenacity}) is the natural
first response, but it lacks centralised coordination.  When
10~agents each independently retry after a 429 error, the
retries arrive simultaneously---the ``thundering
herd''~\cite{dean2013tail}---re-triggering the rate limit.
\hm{}'s centralised retry serialises retries through the
admission gate, preventing amplification.

\subsection{Real-World Validation}

We validated \hm{} against two local model servers
(Table~\ref{tab:realworld}): Ollama~\cite{ollama} serving
Qwen~3.5-4B (GGUF, Q4\_K\_M) and an MLX inference server
serving Qwen~3.5-4B-4bit.  Each test used 10~agents making
3~turns each, with the \texttt{--compare} flag running direct
mode first, then \hm{} mode.

\begin{table}[!htb]
\centering
\small
\caption{Real-world validation against local model servers
(10~agents $\times$ 3~turns).}
\label{tab:realworld}
\begin{tabular}{@{}llcrr@{}}
\toprule
Server & Mode & Alive & Fail\% & Time \\
\midrule
Ollama & Direct  & 10/10 & 0\% & 30.5\,s \\
Ollama & \hm{}   & 10/10 & 0\% & 28.5\,s \\
\midrule
MLX    & Direct  & 10/10 & 0\% & 3.9\,s \\
MLX    & \hm{}   & 10/10 & 0\% & 3.6\,s \\
\bottomrule
\end{tabular}
\end{table}

Local models handle concurrency gracefully (they queue
internally), so these tests do not trigger the stampede scenario.
They do, however, confirm that \hm{} adds negligible overhead:
$<$3\,ms per proxied request, and in the Ollama case, \hm{} was
actually 7\%~faster than direct access because its admission gate
($C_{\max} = 2$) matched Ollama's natural concurrency and reduced
internal queuing contention.

An earlier test run produced one MLX failure (10\%) caused by a
\texttt{RemoteProtocolError} (``server disconnected'').  Adding
this pattern to the retryable-error list
(Section~\ref{sec:retry}) resolved the issue; subsequent runs
achieve 10/10 across both servers.

\begin{figure}[t]
\centering
\includegraphics[width=\columnwidth]{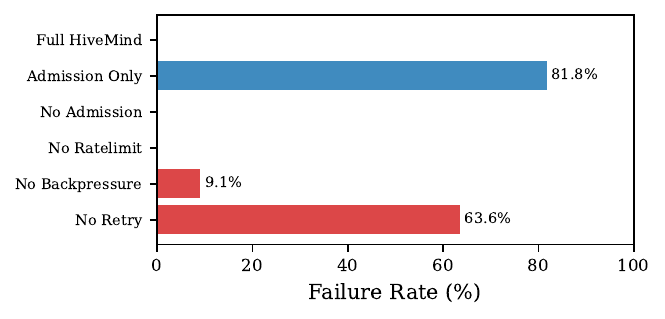}
\caption{Ablation study results.  Removing retry causes the
largest degradation (63.6\% failure); admission-only is
insufficient (81.8\%).  Other primitives are compensated by
the remaining ones.}
\label{fig:ablation}
\end{figure}

\subsection{Cost of Wasted Compute}

Token waste translates directly to monetary cost.
Table~\ref{tab:cost} shows the cost of wasted tokens (tokens
consumed by agents that ultimately failed) across our evaluation
suite, extrapolated to a daily workload of 10~runs.

\begin{table}[!htb]
\centering
\footnotesize
\caption{Daily cost of wasted tokens at current Anthropic
pricing (per million input tokens), assuming 10~evaluation runs
per day.}
\label{tab:cost}
\begin{tabular}{@{}l rr r@{}}
\toprule
Model & Direct & \hm{} & Savings \\
\midrule
Haiku (\$0.80/M)  & \$0.35 & \$0.01 & 97\% \\
Sonnet (\$3/M)    & \$1.31 & \$0.05 & 96\% \\
Opus (\$15/M)     & \$6.55 & \$0.24 & 96\% \\
\bottomrule
\end{tabular}
\end{table}

\noindent At Opus-tier pricing, uncoordinated agents waste
\$6.55/day from our seven-scenario suite alone.  In production
workloads with 20--50 agents running continuously, waste scales
to hundreds of dollars per day.  \hm{} reduces this by 96--97\%.

\begin{figure}[t]
\centering
\includegraphics[width=0.85\columnwidth]{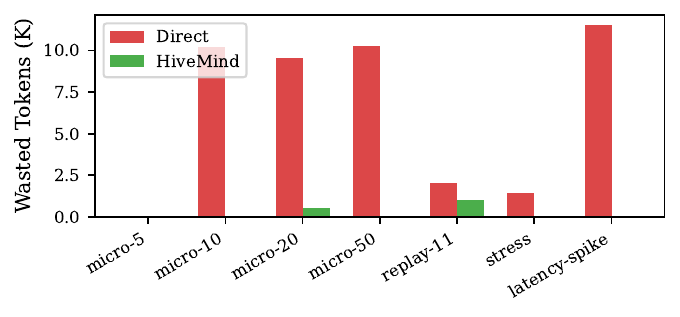}
\caption{Wasted tokens by scenario (thousands).  Direct mode
(red) wastes 1--12K tokens per scenario; \hm{} (green)
reduces waste to near-zero.}
\label{fig:waste}
\end{figure}

% ============================================================
\section{Related Work}
\label{sec:related}

\paragraph{Agent orchestration frameworks.}
LangChain~\cite{langchain2022}, CrewAI~\cite{crewai2024},
AutoGen~\cite{autogen2023}, and Semantic
Kernel~\cite{semantickernel2023} focus on agent composition---chains,
crews, multi-agent conversations---but not on resource
management.  They assume the API is always available and
delegate retry to per-request libraries.  \hm{} is
complementary: it sits below any of these frameworks, managing
the shared API resource that they all depend on.

\paragraph{API rate-limiting libraries.}
Libraries like \texttt{tenacity} and \texttt{backoff} provide
retry logic at the individual request level but lack system-wide
coordination.  Each agent retries independently, potentially
amplifying load during rate-limit windows---the ``thundering
herd'' problem~\cite{dean2013tail}.  \hm{} centralises retry
decisions across all agents sharing an API key.

\paragraph{TCP congestion control.}
Our AIMD backpressure controller directly adapts the Additive
Increase / Multiplicative Decrease algorithm from TCP
Tahoe/Reno~\cite{jacobson1988congestion,chiu1989aimd,allman2009tcp}.
The key insight is that API latency serves the same role as
network round-trip time: it signals congestion before requests
are dropped.  Unlike TCP, we do not implement slow start
(APIs have known baseline concurrency) or fast recovery
(the concurrency space is too small for multiplicative probing).

\paragraph{Circuit breaker pattern.}
Nygard~\cite{nygard2018release} introduced the circuit breaker
as a stability pattern for distributed systems.  Our circuit
breaker adapts this for the LLM API context: the error-rate
threshold is tuned for API-level failures (429, 502), the
half-open probe uses a real API request rather than a health
check, and the state is co-located with the AIMD controller
so that circuit events also trigger concurrency reduction.

\paragraph{Staged event-driven architecture.}
Welsh et al.'s SEDA~\cite{welsh2001seda} proposed decomposing
Internet services into stages connected by queues, with each
stage applying admission control independently.  \hm{}'s
pipeline (admission $\to$ rate limit $\to$ backpressure $\to$
forward $\to$ retry) follows the same staged pattern, though
our stages are co-located in a single process rather than
distributed across threads.

\paragraph{OS scheduling theory.}
The correspondence between LLM agent scheduling and process
scheduling has not been previously formalised in the literature.
Classical scheduling theory---shortest-job-first, priority
scheduling, multilevel feedback
queues~\cite{silberschatz2018os,tanenbaum2015os}---applies
directly when the ``CPU'' is an API request slot and
``processes'' are stateful agents with unpredictable execution
times.  The concurrency primitives
(semaphores~\cite{dijkstra1965cooperating}, condition variables,
monitors~\cite{herlihy2012art}) translate directly to the
async-I/O domain.

\paragraph{SWE-bench and agent reliability.}
SWE-bench~\cite{jimenez2024swebench} and
SWE-agent~\cite{yang2024sweagent} evaluate agent success rates
on real GitHub issues but do not isolate API-access failures as
a distinct cause of task failure.  Our work addresses a failure
mode that is orthogonal to agent capability: an agent may be
perfectly capable of solving a task but still fail because its
API call was dropped by the provider.

% ============================================================
\section{Discussion}
\label{sec:discussion}

\subsection{Design Tradeoffs}

\paragraph{Proxy vs.\ SDK integration.}
We chose a transparent HTTP proxy over SDK-level integration
(e.g., a custom \texttt{httpx} transport or a LangChain
callback).  This sacrifices per-request metadata (the proxy
cannot read agent-internal state) but gains universality: the
same proxy works for Python, TypeScript, Go, and shell-based
agents without any code changes.  The MCP server mode provides
richer integration for agents that support tool use.

\paragraph{Condition variable vs.\ semaphore.}
The condition-variable admission gate adds ${\sim}$50\,$\mu$s of
overhead per acquire/release compared to a raw semaphore.  This
is negligible relative to API latency (typically 500--5000\,ms)
and eliminates undefined behaviour during dynamic resizing.

\paragraph{AIMD tuning.}
The default AIMD parameters ($\alpha = 0.5$, $\beta = 0.5$,
$L_{\text{target}} = 2000$\,ms) are conservative.  Provider
profiles override these: Ollama uses $\beta = 0.7$ (gentler
decrease, since local inference doesn't benefit from aggressive
backoff) and $L_{\text{target}} = 10\,000$\,ms (local models
are inherently slower).  These defaults can be further tuned
via the \tool{hm.config} tool at runtime.

\paragraph{Circuit breaker placement.}
The circuit breaker is co-located with the AIMD controller
rather than implemented as a separate middleware layer.  This
ensures that circuit-open events also reduce the AIMD
concurrency level, preventing a burst of requests when the
circuit closes.

\subsection{Limitations}

\begin{itemize}[nosep]
  \item \textbf{Single-machine scope.}  The current
    implementation runs on one machine.  Distributed scheduling
    across multiple machines sharing an API key is architecturally
    supported via Redis-backed state but not yet evaluated at
    scale.
  \item \textbf{Token estimation.}  When provider tokenizers
    are unavailable, token counting uses a heuristic
    (4~chars/token).  This underestimates for languages with
    long tokens (e.g., CJK) and overestimates for code with
    short identifiers.
  \item \textbf{Mock evaluation.}  Our primary evaluation uses
    a mock API server supporting both Anthropic and OpenAI
    response formats.  While it simulates realistic behaviour
    (rate limits, errors, latency, streaming), the stampede
    failure mode requires a cloud API with hard rate limits to
    trigger reliably---local models queue gracefully.
  \item \textbf{Dynamic priority.}  Priority is set at submission
    time.  Automatic priority adjustment based on observed
    progress (e.g., promoting agents near completion) is future
    work.
  \item \textbf{No cross-agent coordination.}  \hm{} manages
    API access but does not coordinate agents' filesystem
    operations or tool calls.  Two agents writing to the same
    file remain a user-level concern.
\end{itemize}

\subsection{Future Directions}

Production-scale validation against cloud APIs (Anthropic,
OpenAI) with 20--50 concurrent agents would strengthen the
empirical claims.  Integrating provider-specific tokenizers
would improve budget accuracy.  A multilevel feedback queue
(promoting short-running agents, demoting long-running ones)
could improve average completion time.  Finally, combining
\hm{} with task-level resilience systems (checkpointing,
decomposition) would provide end-to-end fault tolerance from
the API layer to the agent layer.

% ============================================================
\section{Conclusion}
\label{sec:conclusion}

We have presented \hm{}, a scheduling system that applies
OS~scheduling principles to concurrent LLM agent workloads.
The five scheduling primitives---admission control via condition
variables, provider-aware rate-limit tracking, AIMD backpressure
with circuit breaking, per-agent token budgets, and priority
queuing with dependency DAGs---in combination eliminate the
failure modes that currently plague parallel agent execution.
The transparent proxy architecture requires zero changes to
existing agents, making \hm{} a drop-in improvement for any
multi-agent workflow.

Our evaluation across seven scenarios shows that uncoordinated
agents fail at 72--100\% rates under contention, while \hm{}
reduces failures to 0--18\%.  An ablation study yields a key
insight for the field: in the current API landscape,
\emph{transparent centralised retry} is more important than
\emph{admission control} for agent survival, but both are most
effective in combination.  This suggests that LLM agent
orchestration systems should prioritise retry coordination over
simple concurrency limiting.

Real-world validation against local model servers confirms that
\hm{} adds under 3\,ms of proxy overhead per request.
Auto-detected provider profiles ensure that the system is
correctly tuned for each API provider out of the box.

A 174-test suite validates correctness across all scheduling
primitives, and the system is open-source under the MIT license
at \url{https://github.com/jayluxferro/hivemind}.

\balance
\begingroup
\sloppy
\bibliographystyle{ieeetr}
\bibliography{references}
\endgroup

\end{document}